# Enhanced Intra Prediction for Video Coding by Using Multiple Neural Networks

Heming Sun, *Member, IEEE*, Zhengxue Cheng, *Student Member, IEEE*, Masaru Takeuchi, *Member, IEEE*, Jiro Katto, *Member, IEEE*

*Abstract*—This paper enhances the intra prediction by using multiple neural network modes (NM). Each NM serves as an end-to-end mapping from the neighboring reference blocks to the current coding block. For the provided NMs, we present two schemes (appending and substitution) to integrate the NMs with the traditional modes (TM) defined in high efficiency video coding (HEVC). For the appending scheme, each NM is corresponding to a certain range of TMs. The categorization of TMs is based on the expected prediction errors. After determining the relevant TMs for each NM, we present a probability-aware mode signaling scheme. The NMs with higher probabilities to be the best mode are signaled with fewer bits. For the substitution scheme, we propose to replace the highest and lowest probable TMs. New most probable mode (MPM) generation method is also employed when substituting the lowest probable TMs. Experimental results demonstrate that using multiple NMs will improve the coding efficiency apparently compared with the single NM. Specifically, proposed appending scheme with seven NMs can save 2.6%, 3.8%, 3.1% BD-rate for Y, U, V components compared with using single NM in the state-of-the-art works.

*Index Terms*—High Efficiency Video Coding (HEVC), intra prediction, neural network, probability

## I. INTRODUCTION

WITH the rapid development of the multimedia society, video traffic is forecasted to occupy more than 80% internet traffic by 2021 as described in [1]. Therefore, video compression technology becomes quite important. Intra prediction is an essential component in the video compression, which can reduce the spatial redundancy by utilizing the similarity of the adjacent pixels. For each coding block, the predicted pixels are the linear regression of the neighboring reference pixels. The prediction is performed according to the mode and size. The available prediction modes have been increased in the past few video compression standards. In H.264/AVC [2], two non-directional modes (Planar, DC) and eight directional modes are supported, while for the successor HEVC [3], the amount of directional modes grows from 8 to 33. To further enhance the prediction accuracy, the number of directional modes is expanded to 65 for the next-generation standard versatile video coding (VVC) as reported in [4]. Regarding the prediction size, it is from 4×4 to 16×16 in H.264, while the maximum size is enlarged to 64×64 in HEVC. In VVC, in addition to the square shape, intra sub partition (ISP) is also adopted to support non-square prediction sizes [5]. With more possible combinations of prediction modes and sizes, the prediction error can be decreased thus the coding efficiency can be improved.

Apart from the expansions of the prediction mode and size, there have also been many other developments to ameliorate the prediction accuracy of the intra prediction. Considering that the correlation between the reference pixels and coding block pixels will become higher with shorter distance, some in-block reconstruction methods are presented. Abdoli *et al*. [6] performed in-block reconstruction for each 4×4 block at the pixel level. Similarly, when ISP is exploited, the reconstruction is sub-block level for each block. To capture the subtle directional texture inside the block, Markov model-based filtering schemes have also been proposed to substitute the traditional copying-based predictions. Chen *et al*. [7] and Li *et al*. [8] proposed a recursive extrapolation with a 3-tap and 4-tap filters. Chen *et al*. [9] presented an iterative filtering method to smooth the predicted pixels. Different from the aforementioned methods, Yeh *et al*. [10] presented a bi-directional intra prediction based on two neighboring predictors. Image inpainting methods were also used for the intra prediction as reported in [11]-[12]. Li *et al*. [13]-[14] utilized multiple lines rather than a single line as the reference pixels. Zhang et al. [15] proposed a hybrid prediction by using both local and non-local correlations.

Recently, there have been several standard works for the VVC intra prediction. Linear regression models are presented to reduce the redundancy between luma and chroma components as described in [16]-[17]. Both non-filtered and filtered reference samples are used for the prediction based on the position information as proposed in [18]-[19]. Besides, four-tap intra interpolation filters are utilized to improve the directional intra prediction accuracy as reported in [20], and

Manuscript received May 8, 2019; revised November 12, 2019; accepted December 12, 2019. This work was supported in part by JST, PRESTO Grant Number JPMJPR19M5, Japan and a research fund from Fujitsu. (*Corresponding author: Zhengxue Cheng*)

H. Sun is with the Waseda Research Institute for Science and Engineering, Tokyo 169-8555, Japan and JST, PRESTO, 4-1-8 Honcho, Kawaguchi, Saitama, 332-0012, Japan (e-mail: hemingsun@aoni.waseda.jp).

Z. Cheng and J. Katto are with the Graduate School of Fundamental Science and Engineering, Waseda University, Tokyo 169-8555, Japan (e-mail: zxcheng@asagi.waseda.jp, katto@waseda.jp).

M. Takeuchi is with the Waseda Research Institute for Science and Engineering, Tokyo 169-8555 (e-mail: masaru-t@aoni.waseda.jp)

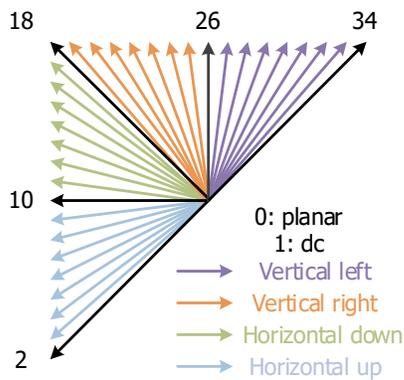

Fig. 1 35 HEVC intra modes.

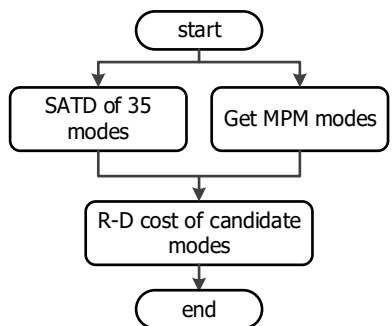

Fig. 2 The procedure of the best luma mode selection in HEVC Test Model.

multiple reference lines are also adopted as stated in [21].

Although the aforementioned literatures have promoted the intra prediction, they are all based on the linear calculation. In fact, recently, neural networks have shown its strong ability in image and video compression due to its powerful non-linear mapping ability. The related works will be introduced in Section II.B. For the intra prediction, there have been also several neural network-based methods. Li *et al.* [22]-[23] exploited fully connected (FC) networks to perform the prediction by using the multiple neighboring pixels. After obtaining the pre-trained model, each model can be regarded as a neural network mode (NM). By integrating these NMs with traditional modes (TM) defined in HEVC, up to 7.4% bits can be saved as reported in [23]. Pfaff *et al.* [24] applied at most 35 NMs on non-square blocks and employed a neural network to select the most probable NM. On average, about 3% BD-rate can be saved compared with original hevc test model (HM).

Though the previous works have achieved significant coding gains by using neural networks, there are still several opportunities for the further improvements. First is to use multiple NMs. For [23], the author exploited two NMs and shown that dual NMs can achieve better coding efficiency than single NM. For [24], the author did consider multiple NMs. However, only one NM is finally selected for the prediction. With more NMs, the prediction accuracy ought to be enhanced which can bring less distortions. On the other hand, more NMs spend larger mode signaling which will increase the bit count. Therefore, an optimal trade-off between rate and distortion of using multiple NMs is expected. Secondly, the authors in [22]-[24] append NMs to TMs which will increase the number of overall modes. As a result, extra flags are required to signal the additional modes. Therefore, maintaining the total number of modes also worth attempting.

In this paper, we focus on the prediction of multiple NMs starting from a fixed block size 8×8. Detailed analysis are performed to show the effect of using multiple NMs. The main contributions are as follows.

1) Two schemes (appending and substitution) to integrate NMs with TMs: We present two integration schemes and analyze their coding gains and complexities thoroughly. For the appending scheme, NMs are regarded as additional modes to TMs. For the substitution schemes, TMs are replaced by NMs.

2) Probability-based TM categorization and mode signaling for the appending scheme: When appending multiple NMs, certain TMs are clustered to correspond with one NM based on the expected prediction error. Besides, a mode signaling scheme is performed for each NM based on the probability.

3) Probability-based TM selection for the substitution scheme: We propose to substitute TMs from two aspects of view. One is to replace the highest probable TMs, while the other is to supplant the lowest probable TMs. When replacing the lowest probable TMs, we present the most probable mode (MPM) generation scheme.

Experimental results demonstrate that using multiple NMs can achieve significant BD-rate saving. Compared with using single NM, 2.6%, 3.8%, 3.1% BD-rate are saved for three channels, respectively.

The rest of the paper is organized as follows. Section II describes the related work. Section III and IV presents our proposals and the training process, respectively. The results are demonstrated in Section V, followed by the conclusions and future work in Section 0.

## II. RELATED WORK

This section is composed of two parts. First, we introduce the intra prediction in HEVC. Second, we review some recent neural network-based methods in image/video compression.

### A. HEVC Intra Prediction

There are 35 intra modes supported in HEVC as shown in Fig. 1. The modes can be categorized to non-directional and directional modes. The former is composed of two modes: Planar and DC, and the latter can be classified according to their directions. Among all the 35 modes, the best luma mode decision is conducted with two steps. In the first step, a list of candidate modes is constructed, and then the final best mode is decided from the candidate mode list in the second step. The overall process is shown in Fig. 2. From the 35 modes, eight modes are selected based on sum of absolute transformed differences (SATD) cost. After that, at most two modes from the MPM set are appended in the candidate list. The MPM set is constructed based on the mode information of the left and above units. Among all the candidate modes, the best mode is selected by comparing rate-distortion (R-D) cost.

When calculating rate for each candidate modes, it is composed of the bits for the mode signaling and the bits for the residual. Several syntax elements (SE) are provided for mode signaling as shown in Fig. 3. For the luma mode, the SE *prev_intra_luma_pred_flag* is used to indicate whether the best mode is from MPM set or not. If the mode is from MPM set, it is binarized as 1. Otherwise, it is binarized as 0. If the best mode belongs to the MPM set, *mpm_idx* specifies the index of three MPM modes. Two *bins* are utilized to distinguish three indexes. If the best mode does not belong to the MPM set, *rem_intra_luma_pred_mode* is used to indicate the number of the remaining modes. Five *bins* are used to represent 32 cases. For the chroma mode, *intra_chroma_pred_mode* is used to indicate the index of five candidate modes. Three *bins* are used to indicate the index.

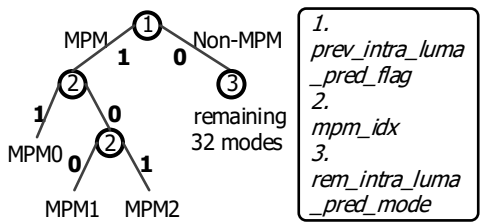

Fig. 3 Mode signaling for 35 luma modes. *prev_intra_luma_pred_flag* represents whether the best mode is from MPM set or not. *mpm_idx* represents the index of 3 MPM modes. *rem_intra_luma_pred_mode* represents the index of 32 non-MPM modes.

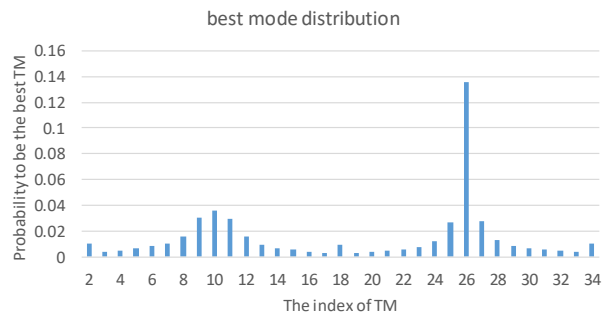

TABLE I
PROBABILITIES OF NON-DIRECTIONAL AND DIRECTIONAL TMS FOR BEING THE BEST MODE

| TM | Prob. (%) |
|---|---|
| Non-directional | 51.1 |
| Directional | 48.9 |

Fig. 4 The probability distribution of directional TMs being the best mode.

After finishing the luma prediction, the best luma mode is derived for the chroma component. In addition to the derived mode, four modes are added in the candidate mode for chroma. If the best luma mode is not among Planar, DC, TM26 (TM-VER) and TM10 (TM-HOR), these four modes and the derived luma mode constitute the candidate mode list. Otherwise, these four modes and TM34 build up the candidate list.

### B. Neural Network-based Methods for Image/Video Coding

There are two types of methods. The first kind is full neural network-based schemes. Toderici *et al*. [25]-[26] exploited multiple layers of recurrent neural networks to recursively compress the residuals. Balle *et al*. [27]-[28] utilized convolutional neural networks (CNN) to construct two autoencoders among which one is used to estimate the coded bit range for the other. Cheng *et al*. [29] analyzed the energy compaction property of learned image compression for non-linear neural network systems. Those works have outperformed the latest image compression knows as HEVC intra prediction (i.e. BPG) in terms of MS-SSIM. Regarding to the video compression, Rippel *et al*. [30] presented a learned end-to-end framework that is better than all the existing standards concerning the quality matrix of MS-SSIM. The second type is adopting neural network technology in the components of traditional compression standards. For the intra prediction, [22]-[24] treated neural networks as new prediction modes to enhance the prediction accuracy. For the inter prediction, [31]-[35] improved the motion interpolation accuracy by using CNN. Regarding the loop filters and post-processing, quite a few of works [36]-[40] have been reported and some CNN-based filters have been proposed with respect to the VVC. About discrete cosine transform (DCT), there have also been several literatures [41]-[42] using CNN to replace the traditional transforms. Xu *et al*. [43] used deep learning for the video transcoding. Lin *et al*. [44] improved the coding gain at low-bitrates by learning-based super resolution and adaptive block patching.

### III. PROPOSED INTRA CODING BASED ON MULTIPLE NEURAL NETWORK MODES

In this section, we will formulate the problem at first, and then present two methods to integrate the NM with TM.

#### A. Mode Probability Analysis and Problem Formulation

In the real coding, the probabilities of 35 TMs to be best mode are different. We encode 2550 still images provided in the New York city library [49] with a moderate quantization parameter (QP) 27. The probabilities of non-directional and directional TMs being the best mode are given in Table I, and the probabilities of all the directional TMs being the best mode are demonstrated in Fig. 4. From the results, we can see that the non-directional TMs have higher probability than directional TMs to be the best mode. For the horizontal TMs from 2 to 18, the distribution is symmetric to TM-HOR. The probability of TM-HOR is the highest and decreases when the distance with TM-HOR becomes larger. Noted that for TM2 and TM18, the probability becomes larger than the neighboring TMs. For the vertical TMs from 18 to 34, the distribution is similar with that for the horizontal TMs.

When using multiple NMs, each NM is corresponding with certain TMs and there are a wide variety of the correspondence. Given that we have N NMs, and the relevant TM set for NMs is marked as S. The target is to minimize the prediction error compared with TM, which can be formulated as the following problem

$$\min_{N,S} \Delta D = \min_{N,S} \left( \sum_{i=0}^{N-1} P_i * \Delta D_{S(i)} \right)$$

$$= \min_{N,S} \left( \sum_{i=0}^{N-1} P_i * \left( D_{S(i)}^N - D_{S(i)}^T \right) \right)$$

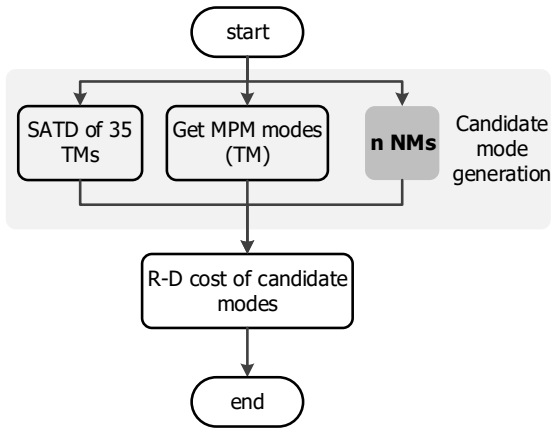

Fig. 5 Proposed luma mode selection for the appending schemes. n (1,3,5,7) is the number of appending NMs.

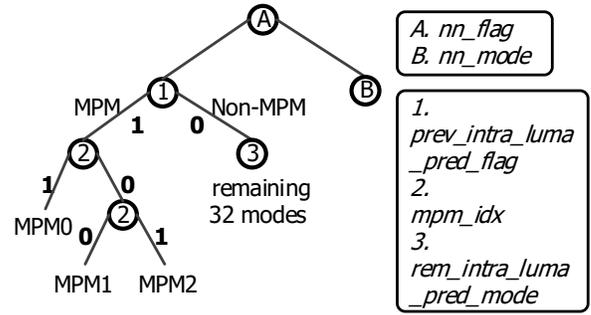

Fig. 6 Proposed mode signaling for (35+n) luma modes.

$$= \min_{N,S} \left( \sum_{i=0}^{N-1} \left( \sum_{j \in s(i)} p_j \right) * \left( D_{s(i)}^N - \frac{\sum_{j \in s(i)} d_j^T}{|s(i)|} \right) \right) \quad (1)$$

where $s(i)$ is the TM set for the i-th NM, $D_{s(i)}^N$ and $D_{s(i)}^T$ are the MSE for the i-th NM and average MSE for the i-th TM set, $P_i$ is the probability of the i-th NM, which can be calculated by the summation of the probabilities of the TMs in $s(i)$, $p_j$ is the probability of the j-th TM.

*B. Appending Neural Network Modes to Traditional HEVC Modes*

When appending NMs, the purpose is to use several NMs to achieve better prediction accuracy than 35 TMs. Therefore, the TM sets of all the NMs should satisfy the following equation

$$\bigcup_{i=0}^{N-1} s(i) = [0,34] \quad (2)$$

where N is the number of supported NMs, and s(i) is the TM set for the i-th NM. In this section, we propose the methods for N being one, three, five and seven.

When appending N NMs, overall (35+N) modes are supported, and the proposed method to select the best luma mode is shown in Fig. 5. In addition to the TMs selected by the SATD cost and the MPM set, NM is also included in the candidate list. Noted that NM is not involved in the MPM generation. Therefore, for each block, if the best mode is NM, we have to set a TM for the sake of MPM generation of its right and bottom blocks. After generating the candidates, R-D cost is calculated to decide the best mode.

If one NM is provided, according to Eq. (2), this NM is corresponding to all the 35 TMs, thus we have entire 36 modes. For the binarization of these 36 modes, a mode signaling scheme is adopted as shown in Fig. 6. One extra SE *nn_flag* is used to distinguish whether the best mode is NM or not. If the best mode is NM, *nn_flag* is set as 1. Otherwise, *nn_flag* is assigned as 0. In the case of the best mode being TM, the mode signaling for the 35 modes maintains the same as the original HEVC, as described in Section II.A. By comparing the R-D cost, if the best mode is NM, we set MPM0 as the best TM. It is because MPM0 has the highest probability to be selected as reported in [43].

When appending one NM, $\Delta D$ in Eq. (1) is 23 according to the experimental results. The positive value means that the prediction error of NM is larger than that of TMs. In order to enhance the prediction accuracy, we increase the number of NMs from one to three. First, 35 TMs can be categorized to non-directional and directional TMs. For the directional TMs, due to a good symmetrical property, the horizontal and vertical TMs are grouped into individual category. These three NMs are symbolled as NM3-NA (non-directional), NM3-VER and NM3-HOR as shown in Table II, and the TM set for the three NMs are [0,1], [2,18] and [18,34], respectively.

For the 38 (35+3) modes, the signaling method is presented in Fig. 6. The difference with appending one NM is that extra SE *nn_mode* is required to represent the best mode. The binarization for the three NMs is shown in Table II. For each NM, the probability is estimated by the summation of the probabilities of the oriented TMs. Thus, the probability of NM3-NA is higher than NM3-HOR and NM3-VER. For the NMs with higher probability, few *bins* are assigned. Therefore, one *bin* is assigned for NM3-NA while two *bins* are assigned for the other two NMs. As shown in Table II, b0 represents whether the NM is non-directional or not, and b1 means whether the NM is horizontal or not.

By appending three NMs, the results of $\Delta D_{s(i)} (i \le 2)$ are shown in Table III. When using three NMs, $\Delta D$ can be reduced to 2.62 by Eq. (1). However, positive value indicates that the prediction error of NM is still larger than that of TM. Therefore, we further increase the number of NMs. From the results in Table III, we can observe that the prediction accuracy of non-directional NM has surpassed that of non-directional TMs. To further enhance the accuracy of directional prediction, the number of directional NMs is increased from two to four. Due to the good symmetric property, horizontal and vertical TMs are symmetrically divided into two categories, and the symbols are given in Table II.

TABLE II
BINIRIZATION OF *NN_MODE* FOR APPENDING 1, 3, 5 AND 7 NMs.

| # of NMs | Binirization b0 | b1 | b2 | b3 | Symbol |
|---|---|---|---|---|---|
| 1 | | | | | NM1 |
| 3 | 1 | | | | NM3-NA |
| | 0 | 1 | | | NM3-HOR |
| | 0 | 0 | | | NM3-VER |
| 5 | 1 | | | | NM5-NA |
| | 0 | 1 | 1 | | NM5-HOR0 |
| | 0 | 1 | 0 | | NM5-HOR1 |
| | 0 | 0 | 0 | | NM5-VER0 |
| | 0 | 0 | 1 | | NM5-VER1 |
| 7 | 1 | | | | NM7-NA |
| | 0 | 0 | 1 | 1 | NM7-HOR0 |
| | 0 | 1 | 1 | | NM7-HOR1 |
| | 0 | 0 | 1 | 0 | NM7-HOR2 |
| | 0 | 0 | 0 | 0 | NM7-VER0 |
| | 0 | 1 | 0 | | NM7-VER1 |
| | 0 | 0 | 0 | 1 | NM7-VER2 |

TABLE III
$\Delta D_{s(i)}$ FOR APPENDING ONE, THREE AND FIVE NMS

| | 1NM | 3 NMs | | | 5 NMs | | | | |
|---|---|---|---|---|---|---|---|---|---|
| i | 0 | 0 | 1 | 2 | 0 | 1 | 2 | 3 | 4 |
| $\Delta D_{s(i)}$ | 23 | -10 | 12 | 19 | -10 | -1 | 12 | 9 | -1 |

In order to represent five NMs, three *bins* are required for *nn_mode*. Similar to appending three NMs, only one *bin* (b0) is consumed in the case of NM-NA. For the four directional NMs, two additional *bins* are required. b1 indicates whether the NM is horizontal or not, while b2 indicates whether the NM is close to the diagonal direction or not.

By appending five NMs, the results of $\Delta D_{s(i)}(i \leq 4)$ are shown in Table III. By Eq. (1), $\Delta D$ becomes -2.9. The negative value represents that using five NMs can eventually achieve a better prediction accuracy than only using TMs.

When appending five NMs, $\Delta D$ becomes negative. However, the categorization of the directional TMs is completely symmetrical to TM-HOR and TM-VER. Therefore, the TMs with high probabilities to be the best mode are categorized into different groups. For instance, TM25 is categorized in NM5-VER0, while TM27 is categorized in NM5-VER1. As a consequence, the probabilities of directional NMs are quite close, and the bin count for each directional NM is identical. In order to solve this problem, we propose to append seven NMs. For the horizontal TMs, we supply three NMs that are corresponding to the TMs of $[2,9-\Delta_1]$, $[10-\Delta_1,10+\Delta_1]$ and $[11+\Delta_1,18]$. By doing so, the TMs around TM-HOR can all be clustered into $[10-\Delta_1,10+\Delta_1]$. Similarly, we also distribute the vertical TMs into three categories that are $[18,25-\Delta_2]$, $[26-\Delta_2,26+\Delta_2]$ and $[27+\Delta_2,34]$. $\Delta D$ in Eq. (1) can be calculated as

$$\Delta D = \Delta D_{NA} + \Delta D_{HOR}(\Delta_1) + \Delta D_{VER}(\Delta_2) \quad (3)$$

where $\Delta D_{NA}$, $\Delta D_{HOR}$ and $\Delta D_{VER}$ are for non-directional, horizontal and vertical NMs, respectively. Since the three NMs are correspdoning to different TMs, we minimize their $\Delta D$ individually. For $\Delta D_{NA}$, it is the same as that in appending three and five NMs, which is equal to -5.1. For $\Delta D_{HOR}$ and $\Delta D_{VER}$, they can be calculated by the following equations

$$\Delta D_{HOR}(\Delta_1) = \sum_{i=1}^{3} P_i * \Delta D_{s(i)}$$
$$= \sum_{j \in [2,9-\Delta_1]} p_j * \Delta D_{[2,9-\Delta_1]}$$
$$+ \sum_{j \in [10-\Delta_1,10+\Delta_1]} p_j * \Delta D_{[10-\Delta_1,10+\Delta_1]}$$
$$+ \sum_{j \in [11+\Delta_1,18]} p_j * \Delta D_{[11+\Delta_1,18]} \quad (4)$$

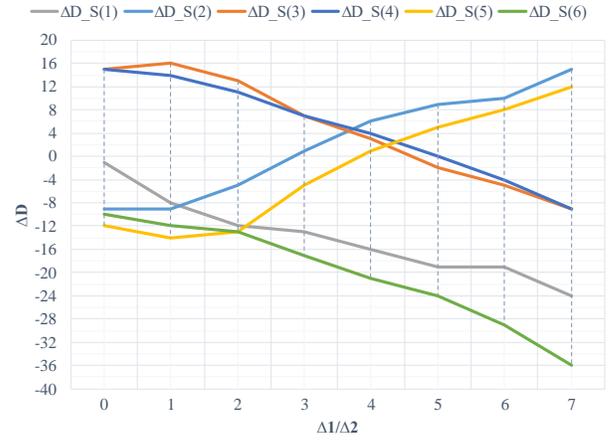

Fig. 7 $\Delta D_{s(i)}$ with various $\Delta 1/\Delta 2$.

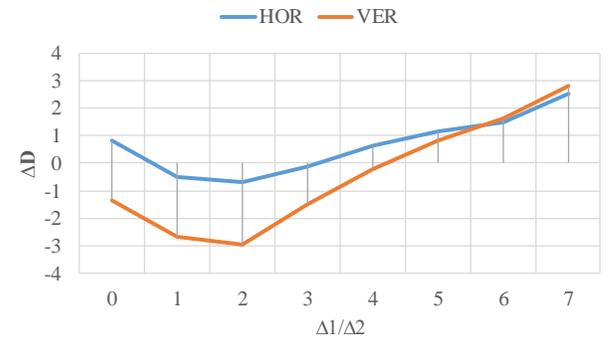

Fig. 8 $\Delta D_{HOR}$ and $\Delta D_{VER}$ with various $\Delta 1/\Delta 2$.

$$\Delta D_{VER}(\Delta_2) = \sum_{i=4}^{6} P_i * \Delta D_{s(i)}$$
$$= \sum_{j \in [18,25-\Delta_2]} p_j * \Delta D_{[18,25-\Delta_2]}$$
$$+ \sum_{j \in [26-\Delta_2,26+\Delta_2]} p_j * \Delta D_{[26-\Delta_2,26+\Delta_2]}$$
$$+ \sum_{j \in [27+\Delta_2,34]} p_j * \Delta D_{[27+\Delta_2,34]} \quad (5)$$

where $\Delta D_{s(i)}(i = 1,2,3)$ are the function of $\Delta_1$. It is because with larger $\Delta_1$, the number of TMs within $[2,9-\Delta_1]$ and $[11+\Delta_1,18]$ is decreased while the number is increased within $[10-\Delta_1,10+\Delta_1]$. Thus, $\Delta D_{s(1)}$ and $\Delta D_{s(3)}$ will be decreased, while $\Delta D_{s(2)}$ will be increased. The results of $\Delta D_{s(i)}(i = 1,2,3)$ under various $\Delta_1$ is shown in Fig. 7. Based on $\Delta D_{s(i)}$, we can calculate $\Delta D_{HOR}(\Delta_1)$ as shown in Fig. 8. We can see that $argmin(\Delta D_{HOR}(\Delta_1))$ is 2 and $min(\Delta D_{HOR}(\Delta_1))$ is -0.7. Similarly, we can find that $argmin(\Delta D_{VER}(\Delta_2))$ is also equal to 2 and $min(\Delta D_{VER}(\Delta_2))$ is -3.0. Overall, $\Delta D$ for appending seven NMs is -8.8, which is smaller than -2.9 when we append five NMs.

For the mode signaling, considering that NM7-NA has the highest probability than the other NMs, only one *bin* is allocated. Same as the mode signaling in appending three and five NMs, b0 is used to indicate whether the best NM is NM7-NA or not. For the six directional NMs, three more *bins* are spent.

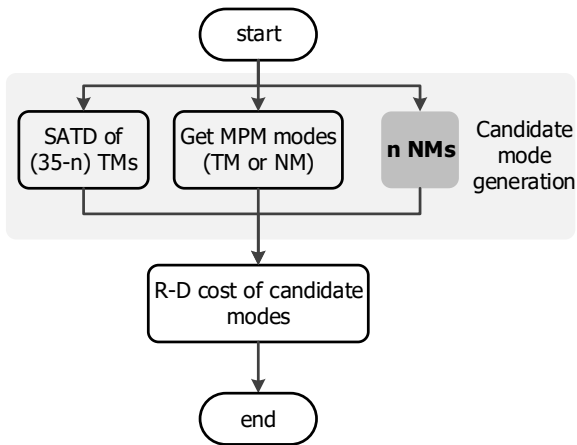

Fig. 9 Proposed luma mode selection for the substituting schemes. n is the number of modes to be replaced.

**Algorithm 1** Proposed MPM generation scheme for substituting low probable TMs
**Input**
best TM of left PU: *PL*
best TM of above PU: *PA*
TM to be substituted: *TMS*
**Output**
3 MPM modes: *M0*, *M1*, *M2*
**if** (*PL==PA*) **then**
　**if** (*PL==TMS*) **then**
　　*M0*=*PL*, *M1* = Planar, *M2* = DC;
　**else if** (*PL==TMS*-1) **then**
　　*M0*=*PL*, *M1* = *PL*-1, *M2* = *PL*+2;
　**else if** (*PL==TMS*+1) **then**
　　*M0*=*PL*, *M1* = *PL*-2, *M2* = *PL*+1;
**else**
　same as origin
**end**

Considering that NM7-HOR1 and NM7-VER1 have higher probability than the other directional NMs, fewer *bins* are assigned. b1 indicates whether the best NM comes from these two NMs. For the other two *bins*, b2 is used to indicate whether the best NM is horizontal NMs while b3 is used to indicate whether the best NM is diagonal.

Similar with that in appending single NM, we have to set a best TM when NM is selected as the best mode. Now that multiple NMs have directional information based on which we can set the related directional TM. When the non-directional NM is selected, it indicates that the current PU does not have explicit directional texture. Therefore, Planar is set as the best TM. If horizontal NMs are selected, it indicates that the current block has more horizontal textures, thus TM-HOR is set as the best TM. Similarly, TM-VER is set as the best TM if the best NM comes from vertical NM set.

After processing the luma component, the next step is to process the chroma component. In the original HM, the best luma mode will be derived as one candidate mode for the chroma prediction. In our proposal, if NM is the best luma mode, in the case of appending one NM and multiple NMs, MPM0 and the best NM will be derived for the chroma prediction, respectively.

## C. Substituting Traditional HEVC Modes by Neural Network Modes

Appending scheme could reduce the bin count when NM is the best mode, while one more *bin* is required when TM is the best mode. In order to avoid the additional bin count, we propose to replace TM(s) by NM(s) to remain the total number of modes.

We propose two kinds of substitution schemes. The first type is focused on replacing high probable TMs, and each NM is corresponding to one TM. Therefore, Eq. (1) is rewritten as

$$\min_{N,S} \Delta D = \min_{N,S} \left( \sum_{i=0}^{N-1} p_{s(i)} * \Delta D_{s(i)} \right) \quad (6)$$

where s(i) is the TM to be replaced. When replacing one TM, Eq. (6) can be simplified as

$$\min_{S} \Delta D = \min_{s(0)} \left( p_{s(0)} * \Delta D_{s(0)} \right) \quad (7)$$

where the range of s(0) is [0,34]. Considering that $\Delta D_{s(0)}$ is minus for all the 35 cases. We can decide that $argmin(\Delta D)$ is 0 according to the probability distribution in Fig. 4, thus Planar is replaced. When substituting three TMs, Eq. (6) is rewritten as

$$\min_{S} \Delta D = \min_{S} \left( p_{s(0)} * \Delta D_{s(0)} + p_{s(1)} * \Delta D_{s(1)} + p_{s(2)} * \Delta D_{s(2)} \right) \quad (8)$$

where $s(i)$ (i=0,1,2) stands for the three TMs that are substituted. According to the distributions in Fig. 4, the three smallest values of $p_{s(i)} * \Delta D_{s(i)}$ appear for i being 0, 1 and 26. Therefore, Planar, DC and TM-VER are substituted.

The second type is focused on substituting low probable TMs. From Fig. 4, we can see that several TMs have very small ratios to be the best mode. Among all the TMs, TM19 has the smallest ratio that is only 0.29%, which means that the contribution of TM19 to the coding gain could be very limited. On the other hand, the NMs such as NM1 probably have better prediction accuracy than TM19 in most scenarios. Therefore, substituting TM19 by NM1 is expected to enhance the coding efficiency. In addition to substituting one TM, we also exploit the effects of replacing three TMs. TM19 is replaced since it is the least significant. In addition, the least significant horizontal-up TM3 and vertical-left TM33 are replaced according to the probability distribution in Fig. 4. NM3-NA, NM3-HOR and NM3-VER are used to replace the three TMs.

The proposed best luma mode selection for the substitution scheme is shown in Fig. 9. Similar with the procedures in the original HM, we will select candidate modes as the first step. The candidate modes are composed of three parts. The first part is selected by comparing the SATD cost. Originally, we have to calculate the SATD costs of all the 35 TMs. Since n TMs are now replaced by NM, thus only the remaining (35-n) TMs require SATD cost computation to pick up eight candidates. The second part of the candidate modes comes from the MPM set. The proposed MPM generation scheme is shown in the pseudo code of Algorithm 1. In the original HM, if the modes of the left and the above PU are the same TM, MPM1 and MPM2 will be set as TM-1 and TM+1, respectively. However, considering that several TM(s) now have been used to represent the NM which does not have the original directional information. Therefore, adding the neighboring directional TM (i.e. TM-1 and TM+1) into MPM set is not reasonable anymore. In our proposed MPM generation method, TM which will be substituted is denoted as TMS. If the best modes of the left and above PU are both TMS, MPM1 and MPM2 are set as Planar and DC, respectively. If the best mode is TMS+1 for both left and above PU, MPM1 is set as TMS in the original HM while TMS-1 in our proposal. If the best mode is TMS-1 for both left and above PU, MPM2 is set as TMS in the original HM while TMS+1 in our proposal. The third part of the candidate modes is the NM. After deciding the candidate modes, the best mode can be selected by comparing the R-D costs.

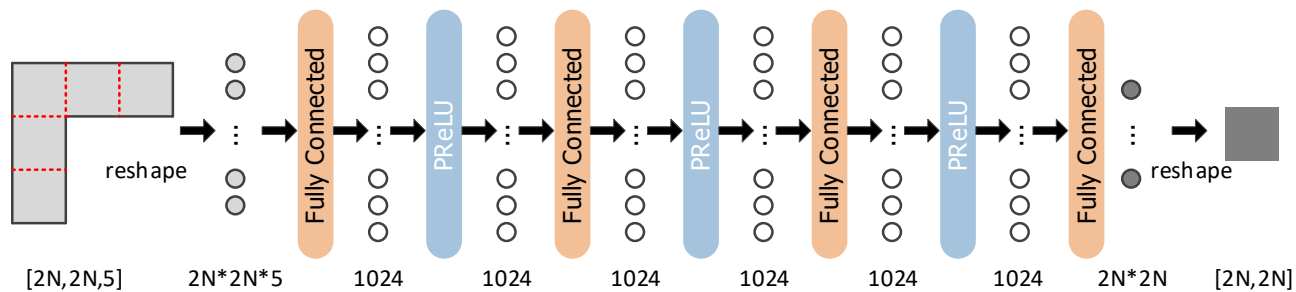

Fig. 10 The network architecture.

For the chroma prediction, the processing method is the same as the original HM. Noted that if the best luma mode is NM, the derived mode for the chroma prediction is also the same NM.

## IV. TRAINING PROCESS

### A. Network Architecture

For each 2N×2N block, the proposed network is shown in Fig. 10. The network is composed of four FC layers and three Parametric Rectified Linear Unit (PReLU) [46] layers. The input/output dimension of each layer is given in Fig. 10. First, multiple pixels of the reference blocks are flattened to one-dimensional vector which is the input of the network. The output of the network is reshaped to 2N×2N size as the predicted pixels for the current block.

About the reference pixels, similar with the previous works [22]-[23], we utilize five neighboring 2N×2N blocks and reshaped to a vector of 2N×2N×5 dimensions as shown in Fig. 10. Take 8×8 as an example, the input is composed of five 8×8 blocks. Therefore, the input dimension of the first FC layer is 320 (8×8×5). For the next three FC layers and PReLU layers, the input dimensions are all 1024. For the final FC layer, the output dimension is 64 (8×8) which can be reshaped to a two-dimensional representation.

The loss function for the training is defined in Eq. (9) where $\Theta$ is the parameter set of weights (W), biases (B) of the FC layer and scale factor of PReLU, $\lambda$ is the weight of the regularization term and M is the batch size. R means the reconstructed pixels of the reference blocks which is the input of the network, and Y represents the original pixels of the current coding block. We will learn the network parameters $\Theta$ to perform the mapping from R to Y.

$$L(\Theta) = \frac{1}{M}\sum_{m=1}^{M} \|F(R^m, \Theta) - Y^m\|_2 + \lambda\|W\|_2^2 \quad (9)$$

$\lambda$ and M are set as 0.0005 and 16 in the training. Weights are initialized as a normal distribution with zero mean and standard deviation of 1. Bias and the scale factor in PReLU are initialized as 0 and 0.25, respectively. ADAM [47] is used for the optimization with the recommended parameters in [47]. The learning rate is set as 0.0001 for the first epoch, and then decayed by 10 and 100 for the next two epochs.

### B. Training Data Generation

We use the New York city library [49] as the training set. All the still images in [49] are encoded by HM version 16.9 [48] to generate the training data. According to [23], the training model has good generalization ability under different quantization parameter (QP) settings. Therefore, we use a moderate QP 27 for encoding all the training sequences. During the encoding, for each 8×8 block, the best coding block will be decided according to the R-D cost. For each best coding block, we fetch the reconstructed pixels of the neighboring five 8×8 blocks. If the pixel has not been reconstructed yet, its closest reconstructed pixel is used to fill the value. The original pixels of the current 8×8 block are fetched as the ground truth. It is noted that there is one difference in building the training set with [22]-[23]. In the previous literatures, the blocks with extremely high complex textures are excluded from the training set. For each block, SATD and mean squared error (MSE) were calculated in [22] and [23] respectively to justify whether it is high complex or not. However, according to our observation, excluding extremely complex blocks will not enhance the prediction accuracy, and calculating SATD or MSE takes time during the training process. Therefore, we include all the blocks in the training set. Overall, 53171363 samples constitute the training set and it takes around eight hours for training one epoch by using GeForce GTX 1080.

## V. EXPERIMENTAL RESULTS

We integrate the proposed method into HM 16.9 [48]. The first frame of 20 sequences in [50] and eight 4K sequences in [51] are encoded by the configuration of "All-intra-main". The proposed method and HM only allow 8×8 intra coding. The BD-rate and BD-psnr are measured for QPs of 22, 27, 32 and

TABLE IV
CODING GAIN OF THE PROPOSED APPENDING SCHEMES.

| Class | Sequence | App1 | | | App3 | | | App5 | | | App7 | | |
|---|---|---|---|---|---|---|---|---|---|---|---|---|---|
| | | Y | U | V | Y | U | V | Y | U | V | Y | U | V |
| A1 (4K) | Tango | -15.6 | -21.5 | -17.7 | -16.9 | -24.7 | -19.2 | -17.4 | -25.0 | -19.3 | -18.4 | -24.7 | -20.8 |
| | Drums100 | -3.7 | -5.0 | -5.7 | -6.1 | -6.4 | -7.3 | -5.0 | -4.5 | -6.9 | -6.7 | -6.1 | -7.6 |
| | CampfireParty | -3.6 | -10.3 | -12.9 | -7.4 | -9.4 | -14.6 | -6.7 | -9.7 | -14.0 | -7.7 | -10.9 | -14.5 |
| | ToddlerFountain | -2.8 | -4.8 | -2.1 | -4.8 | -11.2 | -5.6 | -5.0 | -10.5 | -4.6 | -5.7 | -12.4 | -4.9 |
| **Average of Class A1** | | **-6.4** | **-10.4** | **-9.6** | **-8.8** | **-12.9** | **-11.7** | **-8.5** | **-12.4** | **-11.2** | **-9.7** | **-13.5** | **-12.0** |
| A2 (4K) | CatRobot | -5.5 | -9.9 | -8.4 | -7.1 | -10.7 | -12.0 | -6.8 | -8.8 | -12.1 | -8.0 | -9.8 | -12.9 |
| | TrafficFlow | -8.7 | -9.1 | -15.0 | -7.5 | -10.8 | -17.9 | -7.9 | -8.6 | -17.5 | -8.4 | -10.7 | -17.0 |
| | DaylightRoad | -8.3 | -18.7 | -16.1 | -8.5 | -19.6 | -17.9 | -8.3 | -18.2 | -17.5 | -8.5 | -14.0 | -16.7 |
| | Rollercoaster | -14.9 | -19.9 | -13.9 | -17.2 | -25.8 | -18.2 | -16.7 | -24.9 | -17.3 | -17.8 | -25.6 | -17.8 |
| **Average of Class A2** | | **-9.3** | **-14.4** | **-13.3** | **-10.1** | **-16.7** | **-16.5** | **-9.9** | **-15.1** | **-16.1** | **-10.7** | **-15.0** | **-16.1** |
| A (WQXGA) | Traffic | -3.1 | -3.7 | -5.7 | -4.7 | -4.7 | -5.7 | -4.6 | -4.7 | -6.2 | -5.2 | -5.0 | -5.8 |
| | PeopleOnStreet | -3.2 | -2.7 | -5.5 | -4.4 | -3.6 | -7.9 | -5.2 | -3.5 | -8.7 | -5.4 | -2.8 | -9.0 |
| | Nebuta | -1.6 | -1.9 | -1.9 | -3.1 | -2.0 | -2.4 | -2.8 | -1.4 | -1.9 | -3.8 | -2.0 | -2.4 |
| | SteamLocomotive | -2.9 | -7.7 | -3.5 | -4.0 | -10.5 | -8.3 | -4.2 | -10.5 | -10.1 | -4.8 | -10.1 | -10.1 |
| **Average of Class A** | | **-2.7** | **-4.0** | **-4.1** | **-4.0** | **-5.2** | **-6.1** | **-4.2** | **-5.0** | **-6.7** | **-4.8** | **-5.0** | **-6.8** |
| B (1080P) | Kimono | -6.0 | -5.8 | -7.9 | -9.5 | -10.3 | -12.6 | -10.0 | -9.1 | -12.6 | -10.9 | -10.5 | -14.2 |
| | ParkScene | -2.2 | -4.9 | -2.9 | -3.9 | -6.1 | -1.6 | -3.9 | -5.8 | -2.5 | -4.4 | -5.5 | -2.7 |
| | Cactus | -2.3 | -2.9 | -5.8 | -3.6 | -3.6 | -5.1 | -3.9 | -3.2 | -3.9 | -4.3 | -2.8 | -6.3 |
| | BQTerrace | -9.7 | -13.2 | -9.5 | -9.7 | -12.3 | -9.8 | -7.6 | -12.0 | -8.5 | -9.9 | -14.7 | -7.0 |
| | BasketballDrive | -2.4 | -4.8 | -8.3 | -2.8 | 0.3 | -8.4 | -3.1 | -4.3 | -8.7 | -3.2 | -2.9 | -5.2 |
| **Average of Class B** | | **-4.5** | **-6.3** | **-6.8** | **-5.9** | **-6.4** | **-7.5** | **-5.7** | **-6.9** | **-7.3** | **-6.6** | **-7.3** | **-7.1** |
| C (WVGA) | BasketballDrill | 0.0 | -0.8 | -0.7 | -0.7 | 0.3 | -1.6 | 0.8 | 0.5 | -1.4 | -1.6 | -1.9 | -4.3 |
| | BQMall | -1.6 | -1.0 | -3.8 | -3.1 | -1.1 | -4.4 | -3.2 | -1.0 | -3.0 | -3.5 | -1.5 | -4.3 |
| | PartyScene | -1.1 | -2.0 | -1.5 | -2.0 | -2.0 | -1.9 | -2.4 | -1.7 | -1.2 | -2.4 | -2.0 | -0.8 |
| | RaceHorsesC | -1.4 | -1.5 | -1.3 | -2.6 | -1.5 | -1.5 | -2.4 | -1.2 | -0.5 | -3.1 | -2.8 | -1.7 |
| **Average of Class C** | | **-1.0** | **-1.3** | **-1.8** | **-2.1** | **-1.0** | **-2.4** | **-1.8** | **-0.8** | **-1.5** | **-2.6** | **-2.1** | **-2.8** |
| D (WQVGA) | BasketballPass | -1.3 | -3.9 | -3.0 | -2.6 | -4.7 | 1.0 | -1.6 | -4.0 | 1.8 | -2.7 | -1.9 | 1.8 |
| | BQSquare | -0.9 | 2.3 | -3.9 | -1.4 | 3.1 | -6.9 | -1.5 | 4.0 | -7.1 | -1.6 | 2.2 | -4.9 |
| | BlowingBubbles | -1.1 | -1.4 | -0.3 | -1.9 | -1.6 | -0.9 | -2.4 | -1.1 | 0.3 | -2.7 | -2.0 | -1.2 |
| | RaceHorses | -1.3 | -1.1 | -1.6 | -2.5 | -0.9 | -1.3 | -2.2 | 0.0 | -0.5 | -3.1 | -0.5 | -0.3 |
| **Average of Class D** | | **-1.2** | **-1.0** | **-2.2** | **-2.1** | **-1.0** | **-2.0** | **-1.9** | **-0.3** | **-1.4** | **-2.5** | **-0.5** | **-1.1** |
| E (720P) | FourPeople | -4.7 | -5.4 | -7.0 | -5.7 | -4.8 | -10.5 | -5.8 | -6.6 | -5.8 | -6.0 | -0.9 | -7.6 |
| | Johnny | -7.7 | -8.1 | -9.9 | -7.9 | -11.7 | -10.5 | -7.7 | -5.8 | -12.1 | -8.6 | -9.8 | -9.7 |
| | KristenAndSara | -5.8 | -3.2 | -9.5 | -6.3 | -5.2 | -9.2 | -6.1 | -4.4 | -9.7 | -6.7 | -4.3 | -10.8 |
| **Average of Class E** | | **-6.1** | **-5.6** | **-8.8** | **-6.6** | **-7.3** | **-10.1** | **-6.5** | **-5.6** | **-9.2** | **-7.1** | **-5.0** | **-9.4** |
| **Average** | | **-4.4** | **-6.2** | **-6.6** | **-5.6** | **-7.2** | **-7.9** | **-5.5** | **-6.7** | **-7.6** | **-6.3** | **-7.0** | **-7.8** |

37 by using Bjontegaard's method [52]. Noted that there is no overlap between training set and test set.

### A. Coding Results of Proposed Schemes

In Table IV, we give BD-rate of all the appending schemes for the Y, U, V channels compared with the anchor. When appending one NM, 4.4%, 6.2% and 6.6% bitrate can be saved for three channels, respectively. From the results, we can see that we can save BD-rates for all the test sequences. However, the bit saving fluctuates hugely for different sequences. Especially, for some 4K sequences such as *Tango* and *Rollercoaster*, we can save more than 15% BD-rates for some channels. It is because more NMs are selected as the best mode for these sequences. As a result, the number of bits for the luma mode signaling can be significantly reduced. Originally, for each prediction block, at most six *bins* are required for the mode signaling, while in our method, only one *bin* is consumed if NM is selected. However, if few blocks select NMs as the best mode, there is a one-*bin* penalty for signaling TM compared with the anchor. In this case, the coding gain is small. To verify the aforementioned assumption, we check the ratio of NM being the best mode for all the 28 test sequences at QP27 and the results are shown in Fig. 11. We can see that there is a definite relationship between the BD-rate reduction and NM ratio. The largest average YUV BD-rate reduction occurs for the sequence *Tango* (index 1) which owns the highest NM ratio (84.4%). On the contrary, the smallest average YUV BD-rate reduction is achieved for the sequence *BasketballDrill* (index 18) that has the lowest NM ratio (27.6%).

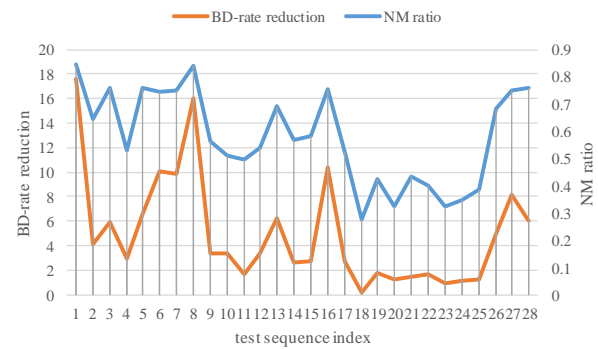

Fig. 11 The relationship between average YUV BD-rate reduction and NM ratio for the 28 test sequences.

TABLE V
CODING GAIN OF THE PROPOSED SUBSTITUTION SCHEMES.

| Class | Sequence | SubL1 | | | SubL3 | | | SubH1 | | | SubH3 | | |
|---|---|---|---|---|---|---|---|---|---|---|---|---|---|
| | | Y | U | V | Y | U | V | Y | U | V | Y | U | V |
| A1 (4K) | Tango | -1.7 | -3.7 | -1.0 | -2.9 | -4.0 | 0.2 | -5.9 | -6.3 | -4.2 | -7.6 | -10.0 | -5.5 |
| | Drums100 | -0.9 | -0.2 | -0.9 | -2.2 | -1.1 | -0.9 | -1.8 | -1.4 | -1.6 | -2.8 | -2.3 | -3.2 |
| | CampfireParty | -1.0 | -4.6 | -4.2 | -1.8 | -5.1 | -3.2 | -2.9 | -3.0 | -3.8 | -2.7 | -2.5 | -5.6 |
| | ToddlerFountain | -1.0 | -1.1 | 0.3 | -1.9 | -2.0 | -1.6 | -1.9 | -2.4 | -1.6 | -3.1 | -4.5 | -2.5 |
| **Average of Class A1** | | **-1.1** | **-2.4** | **-1.5** | **-2.2** | **-3.1** | **-1.4** | **-3.1** | **-3.3** | **-2.8** | **-4.0** | **-4.8** | **-4.2** |
| A2 (4K) | CatRobot | -0.7 | -1.7 | -0.7 | -1.6 | -2.0 | -1.1 | -2.9 | -3.5 | -2.7 | -4.1 | -5.3 | -4.5 |
| | TrafficFlow | -0.9 | -1.0 | -0.3 | -0.3 | -1.5 | 0.7 | -3.2 | -3.7 | -2.4 | -3.5 | -5.7 | -4.5 |
| | DaylightRoad | -1.4 | -4.0 | -0.7 | -2.2 | -6.0 | -2.1 | -3.1 | -5.4 | -3.8 | -4.2 | -8.4 | -5.5 |
| | Rollercoaster | -1.0 | 0.1 | -2.9 | -3.8 | -2.1 | -4.1 | -6.2 | -5.4 | -7.7 | -7.1 | -7.0 | -7.2 |
| **Average of Class A2** | | **-1.0** | **-1.7** | **-1.1** | **-2.0** | **-2.9** | **-1.6** | **-3.8** | **-4.5** | **-4.1** | **-4.7** | **-6.6** | **-5.4** |
| A (WQXGA) | Traffic | -0.8 | -0.7 | -1.5 | -2.0 | -1.0 | -1.2 | -1.9 | -1.9 | -1.1 | -2.5 | -2.5 | -2.7 |
| | PeopleOnStreet | -1.4 | -0.5 | -1.3 | -2.3 | -0.2 | -1.0 | -1.9 | -0.8 | -2.2 | -2.6 | -1.0 | -2.6 |
| | Nebuta | -0.9 | -0.6 | -0.8 | -1.9 | -0.4 | -0.5 | -1.3 | -0.9 | -1.3 | -1.6 | -1.0 | -1.4 |
| | SteamLocomotive | -1.1 | -1.6 | -2.9 | -2.0 | -4.0 | -2.4 | -2.0 | -3.2 | -2.8 | -2.7 | -2.1 | -2.3 |
| **Average of Class A** | | **-1.1** | **-0.8** | **-1.6** | **-2.1** | **-1.4** | **-1.3** | **-1.8** | **-1.7** | **-1.9** | **-2.3** | **-1.6** | **-2.3** |
| B (1080P) | Kimono | -1.5 | -1.2 | -3.1 | -2.9 | -1.6 | -1.6 | -4.1 | -4.0 | -4.2 | -5.9 | -5.8 | -7.6 |
| | ParkScene | -0.7 | 1.2 | -4.6 | -1.5 | -1.3 | -3.2 | -1.7 | -0.7 | 0.1 | -2.1 | -3.3 | -4.6 |
| | Cactus | -0.6 | -0.7 | -0.6 | -1.5 | -0.8 | -1.4 | -1.5 | -1.0 | -2.0 | -1.9 | -1.6 | -2.5 |
| | BQTerrace | 0.3 | 0.5 | -0.9 | -1.1 | -0.5 | -0.7 | -2.4 | -3.8 | -0.5 | -2.9 | -4.6 | -2.5 |
| | BasketballDrive | -0.9 | 0.2 | 2.9 | -1.4 | 0.7 | 1.3 | -1.5 | -1.2 | 1.2 | -2.3 | 1.0 | 0.2 |
| **Average of Class B** | | **-0.7** | **0.0** | **-1.3** | **-1.7** | **-0.7** | **-1.1** | **-2.2** | **-2.1** | **-1.1** | **-3.0** | **-2.9** | **-3.4** |
| C (WVGA) | BasketballDrill | 1.1 | 0.9 | 2.2 | 0.2 | 0.2 | -0.3 | 0.2 | -1.2 | -0.1 | 0.3 | 0.4 | 0.8 |
| | BQMall | -0.9 | 0.7 | -0.8 | -1.5 | 0.4 | -1.5 | -1.3 | 0.7 | -2.2 | -1.9 | -1.3 | -3.1 |
| | PartyScene | -0.7 | -1.1 | -0.1 | -1.1 | -1.3 | -0.6 | -0.9 | -0.6 | -1.3 | -1.2 | -1.1 | -1.3 |
| | RaceHorsesC | -0.6 | -0.2 | -0.5 | -1.3 | -0.4 | 0.1 | -1.1 | -0.5 | 0.3 | -1.2 | -1.1 | -1.3 |
| **Average of Class C** | | **-0.3** | **0.1** | **0.2** | **-0.9** | **-0.3** | **-0.6** | **-0.8** | **-0.4** | **-0.8** | **-1.0** | **-0.8** | **-1.2** |
| D (WQVGA) | BasketballPass | -0.6 | 0.2 | 0.5 | -1.0 | -2.0 | 1.5 | -0.6 | -2.1 | 1.7 | -0.8 | -1.9 | -0.9 |
| | BQSquare | -0.4 | 2.5 | -1.0 | -0.9 | 2.9 | -3.0 | -0.7 | 4.0 | -3.4 | -0.9 | 2.7 | -5.8 |
| | BlowingBubbles | -0.5 | 0.1 | 1.7 | -1.1 | 0.6 | 1.9 | -0.7 | -1.6 | -0.1 | -1.1 | -2.2 | 0.4 |
| | RaceHorses | -0.8 | 0.4 | 0.4 | -1.2 | -0.7 | -1.0 | -1.5 | 0.2 | -0.4 | -1.3 | 0.1 | -0.9 |
| **Average of Class D** | | **-0.6** | **0.8** | **0.4** | **-1.0** | **0.2** | **-0.1** | **-0.9** | **0.1** | **-0.6** | **-1.0** | **-0.3** | **-1.8** |
| E (720P) | FourPeople | -1.0 | 0.7 | 0.3 | -2.5 | 0.1 | -0.6 | -2.8 | -0.5 | -1.4 | -3.7 | -1.3 | -3.3 |
| | Johnny | -0.8 | -1.7 | 0.6 | -1.8 | -5.8 | -2.1 | -4.1 | -3.1 | -2.2 | -5.2 | -5.3 | -6.0 |
| | KristenAndSara | -1.2 | 0.0 | -1.3 | -2.0 | 0.0 | -3.6 | -3.3 | -1.3 | -6.6 | -4.2 | -3.9 | -5.1 |
| **Average of Class E** | | **-1.0** | **-0.3** | **-0.2** | **-2.1** | **-1.9** | **-2.1** | **-3.4** | **-1.7** | **-3.4** | **-4.4** | **-3.5** | **-4.8** |
| **Average** | | **-0.8** | **-0.6** | **-0.8** | **-1.7** | **-1.4** | **-1.2** | **-2.3** | **-2.0** | **-2.0** | **-2.9** | **-2.9** | **-3.3** |

When appending multiple NMs, we can achieve more coding gain. Specifically, 6.3%, 7.0% and 7.8% BD-rates can be saved for three channels when appending seven NMs. We can see that when increasing the number of NMs from three to five, the BD-rate becomes worse, while it becomes better if further increasing the number of NMs to seven. This observation demonstrates that grouping important TMs in one category could enhance the coding efficiency. When appending five NMs, TM9 and TM11 are grouped into two categories. TM25 and TM27 are also located in two categories. As a result, each category has close probability to be selected, which will weaken the efficiency of the context-adaptive coding.

For the substitution scheme, the results of the BD-rates are shown in Table V. When substituting the lowest probable TM (SubL1), less than 1% BD-rate can be saved compared with the anchor. When replacing the three lowest probable TMs (SubL3), 1.7%, 1.4%, 1.2% BD-rates can be saved for the three channels, respectively. When substituting the highest probable TM (SubH1), around 2% BD-rates can be saved compared with the anchor. If the three highest probable TMs are replaced (SubH3), larger BD-rate saving can be achieved for all the three channels. From the results, we can see that replacing high probable TMs is much more efficient than replacing low probable TMs in terms of the coding gain. In the case of substituting one TM, the average Y-BD-rate saving is only 0.8% for SubL1, while it can reach 2.3% for SubH1. In the case of substituting three TMs, the average Y-BD-rate saving is only 1.7% for SubL3, while it can achieve 2.9% for SubH3.

By comparing Table IV with Table V, we can observe that the appending schemes can achieve much better coding efficiency than the substituting schemes. The best coding gain for the substitution scheme is substituting three high probable TMs, which can save 2.9%, 2.9% and 3.3% BD-rates for three channels. However, when appending three TMs, we can save 5.6%, 7.2% and 7.9% BD-rates for three channels, respectively. It is because appending NMs can save the cost of mode signaling compared with the origin. As a result, fewer mode signaling cost can significantly contribute to the coding efficiency in the case of appending NMs. However, since there is no change for the mode signaling for the substitution scheme, thus the coding gains for the substituting schemes mainly come from the prediction accuracy enhancement.

TABLE VI
CODING EFFICIENCY COMPARISON WITH PREVIOUS WORKS

| Class | Sequence | ICIP'17 [22] | | | TIP'18 [23] | | | Proposed App7 | | |
|---|---|---|---|---|---|---|---|---|---|---|
| | | Y | U | V | Y | U | V | Y | U | V |
| A1 (4K) | Tango | -3.3 | -4.8 | -4.5 | -15.6 | -23.0 | -17.4 | -18.4 | -24.7 | -20.8 |
| | Drums100 | -1.5 | -1.9 | -1.6 | -2.5 | 0.6 | -1.5 | -6.7 | -6.1 | -7.6 |
| | CampfireParty | -0.5 | -0.8 | -0.9 | -4.3 | 2.2 | -10.2 | -7.7 | -10.9 | -14.5 |
| | ToddlerFountain | -2.1 | -2.4 | -2.7 | -3.1 | -5.7 | -1.8 | -5.7 | -12.4 | -4.9 |
| **Average of Class A1** | | **-1.9** | **-2.5** | **-2.4** | **-6.4** | **-6.5** | **-7.7** | **-9.6** | **-13.5** | **-12.0** |
| A2 (4K) | CatRobot | -1.0 | -1.7 | -1.3 | -5.7 | -6.7 | -7.9 | -8.0 | -9.8 | -12.9 |
| | TrafficFlow | -1.4 | -1.8 | -1.9 | -7.2 | -7.5 | -16.6 | -8.4 | -10.7 | -17.0 |
| | DaylightRoad | -1.6 | -3.4 | -2.8 | -7.0 | -11.3 | -14.2 | -8.5 | -14.0 | -16.7 |
| | Rollercoaster | -1.8 | -2.5 | -2.1 | -11.6 | -14.3 | -9.3 | -17.8 | -25.6 | -17.8 |
| **Average of Class A2** | | **-1.5** | **-2.4** | **-2.0** | **-7.9** | **-10.0** | **-12.0** | **-10.7** | **-15.0** | **-16.1** |
| A (WQXGA) | Traffic | -1.0 | -1.4 | -1.6 | -1.6 | -0.5 | -1.3 | -5.2 | -5.0 | -5.8 |
| | PeopleOnStreet | -1.3 | -1.6 | -2.0 | -2.9 | -1.1 | -5.8 | -5.4 | -2.8 | -9.0 |
| | Nebuta | -1.6 | -1.4 | -1.5 | -1.4 | 2.1 | 1.3 | -3.8 | -2.0 | -2.4 |
| | SteamLocomotive | -1.7 | -2.3 | -2.6 | -2.9 | -6.8 | -6.3 | -4.8 | -10.1 | -10.1 |
| **Average of Class A** | | **-1.4** | **-1.7** | **-1.9** | **-2.2** | **-1.6** | **-3.0** | **-4.8** | **-5.0** | **-6.8** |
| B (1080P) | Kimono | -3.2 | -4.1 | -4.0 | -5.5 | -4.9 | -8.1 | -10.9 | -10.5 | -14.2 |
| | ParkScene | -1.1 | -1.3 | -1.4 | -2.3 | -1.2 | -1.0 | -4.4 | -5.5 | -2.7 |
| | Cactus | -0.9 | -1.3 | -1.7 | -2.0 | 0.0 | -1.9 | -4.3 | -2.8 | -6.3 |
| | BQTerrace | -0.5 | -2.1 | -1.2 | -4.6 | -13.3 | -2.9 | -9.9 | -14.7 | -7.0 |
| | BasketballDrive | -0.9 | -0.1 | -0.1 | -1.6 | -4.2 | -6.1 | -3.2 | -2.9 | -5.2 |
| **Average of Class B** | | **-1.3** | **-1.8** | **-1.7** | **-3.2** | **-4.7** | **-4.0** | **-6.5** | **-7.3** | **-7.1** |
| C (WVGA) | BasketballDrill | -0.3 | -1.5 | -1.5 | 1.2 | 2.8 | 3.3 | -1.6 | -1.9 | -4.3 |
| | BQMall | -0.3 | -0.3 | -0.5 | -1.4 | 1.6 | 0.2 | -3.5 | -1.5 | -4.3 |
| | PartyScene | -0.4 | -0.5 | -0.4 | -1.1 | 0.2 | 1.5 | -2.4 | -2.0 | -0.8 |
| | RaceHorsesC | -0.8 | -1.5 | -1.1 | -1.2 | 1.7 | 3.4 | -3.1 | -2.8 | -1.7 |
| **Average of Class C** | | **-0.5** | **-1.0** | **-0.9** | **-0.6** | **1.6** | **2.1** | **-2.7** | **-2.1** | **-2.8** |
| D (WQVGA) | BasketballPass | -0.4 | -1.4 | -1.0 | 0.4 | -2.4 | 3.4 | -2.7 | -1.9 | 1.8 |
| | BQSquare | -0.2 | -1.0 | 0.5 | -0.8 | 3.3 | -5.0 | -1.6 | 2.2 | -4.9 |
| | BlowingBubbles | -0.6 | -0.2 | -1.0 | -1.3 | 1.8 | 2.5 | -2.7 | -2.0 | -1.2 |
| | RaceHorses | -0.6 | -1.2 | -1.4 | -1.0 | 2.7 | 3.0 | -3.1 | -0.5 | -0.3 |
| **Average of Class D** | | **-0.5** | **-1.0** | **-0.7** | **-0.7** | **1.4** | **1.0** | **-2.5** | **-0.6** | **-1.2** |
| E (720P) | FourPeople | -0.8 | -1.0 | -2.3 | -3.4 | -0.3 | -6.1 | -6.0 | -0.9 | -7.6 |
| | Johnny | -1.0 | -1.3 | -1.4 | -8.3 | -4.3 | -14.9 | -8.6 | -9.8 | -9.7 |
| | KristenAndSara | -0.8 | -1.1 | -1.1 | -5.8 | -2.6 | -12.7 | -6.7 | -4.3 | -10.8 |
| **Average of Class E** | | **-0.9** | **-1.1** | **-1.6** | **-5.8** | **-2.4** | **-11.2** | **-7.1** | **-5.0** | **-9.4** |
| **Average of All** | | **-1.1** | **-1.6** | **-1.6** | **-3.7** | **-3.2** | **-4.7** | **-6.3** | **-7.0** | **-7.8** |

### B. Coding Efficiency Comparison with Previous Works

Since appending seven NMs can achieve the best coding efficiency among all the proposals, so we compare its coding results with the previous works [22]-[23]. [22] only focused on the 8x8 block prediction, thus we directly use the BD-rate results in the literature. About [23], the author performed a complete comparison based on a variable block size. To compare with [23] and show the effect of proposed multiple NMs, we reimplemented their methods of using one NM and adapted to 8×8, the coding gain compared with the anchor is shown in Table VI. Noted that for the fair comparison, we use the same pre-trained model in our proposal and the reimplementation of [23]. As shown in Table VI, compared with [22], we could achieve significant coding gains. The main reason is that we use more neural network layers and dimensions. In [22], only three layers are adopted and the dimension is 128, while we utilize four layers and the dimension is 1024. It shows that using appropriately more layers and dimensions does improve the network prediction accuracy. Compared with [23], a better coding gain can be achieved due to two differences. The first difference is the TM setting when NM is the best mode. In [23], if NM is the best mode, Planar is set as the best TM for the current block. However, in our proposal, MPM0 rather than Planar is set as

> REPLACE THIS LINE WITH YOUR PAPER IDENTIFICATION NUMBER (DOUBLE-CLICK HERE TO EDIT) <    11TABLE VII
ENCODING AND DECODING COMPLEXITY INCREASE ALONG WITH THE BD-RATES

| Method | Complexity | | BD-rate (%) | | |
|---|---|---|---|---|---|
| | Encoder | Decoder | Y | U | V |
| ICIP'17 [22] | 148% | 290% | -1.1 | -1.6 | -1.6 |
| TIP'18 [23] | 1421% | 13227% | -3.7 | -3.2 | -4.7 |
| App1 | 566% | 3821% | -4.4 | -6.2 | -6.6 |
| App3 | 3149% | 16860% | -5.6 | -7.2 | -7.9 |
| App5 | 4769% | 18171% | -5.5 | -6.7 | -7.6 |
| App7 | 6335% | 19032% | -6.3 | -7.0 | -7.8 |
| SubH1 | 546% | 2316% | -2.3 | -2.0 | -2.0 |
| SubH3 | 2105% | 5828% | -2.9 | -2.9 | -3.3 |
| SubL1 | 748% | 2475% | -0.8 | -0.6 | -0.8 |
| SubL3 | 2410% | 4778% | -1.7 | -1.4 | -1.2 |

the best TM. The second difference lies in the chroma prediction. In [23], if the best luma mode is NM, NM is directly used for the chroma prediction. By doing so, the cost of mode signaling for the chroma components can be reduced while the prediction error will be increased. In our method, NM is just derived as a candidate for the chroma prediction.

On average, compared with [22], we can save 5.2%, 5.4% and 6.2% more BD-rates for Y, U and V channel respectively. Compared with [23], we can save 2.6%, 3.8% and 3.1% more BD-rates for the three channels, respectively.

*C. Coding Complexity Analysis*

In addition to the coding gain, we also give the complexity increasement compared with the anchor in terms of the encoding and decoding time in Table VII. The test is conducted on Intel Core i7-7820X CPU@3.60GHz with 32GB RAM. For the encoding, when appending one NM, the encoding times becomes 5.7x larger. When the number of appending NMs increases, the encoding times becomes 31.5x, 47.7x and 63.4x larger for three, five and seven NMs, respectively. For the substitution scheme, when replacing one and three high probable TM, 5.5x and 21.1x larger encoding time is consumed, respectively. When replacing one and three low probable TM, 7.5x and 24.1x larger encoding time is consumed, respectively. From the results, we can see that replacing low probable TMs consume more encoding time than replacing high probable TMs. It is because when replacing high probable TMs, the NMs are very likely to be included in the MPM set, thus there is no need to additionally add the NMs in the candidate mode list. However, when replacing low probable TMs, the NMs will be included in the MPM set only if the best mode of the neighboring block is the same NM. Therefore, the NMs will be additionally appended in the candidate mode list. As a result, the number of candidate modes of replacing low probable TMs is larger than that of replacing high probable TMs, thus the encoding time is a little bit longer.

For the decoding, when appending one NM, 38.2x larger decoding time is consumed. When appending more NMs, the decoding time becomes 168.6x, 181.7x and 190.3x larger for three, five and seven NMs, respectively. For the substitution scheme, when replacing one and three high probable TMs, 23.2x and 58.3x larger decoding time is consumed, respectively.

When replacing one and three low probable TMs, 24.8x and 47.8x larger decoding time is spent, respectively. The decoding complexity is related with the NM ratios. If NM ratio is higher, the decoding time will be enlarged. Therefore, appending one NM requires larger decoding time than substituting one NM since more NMs are selected when using the appending scheme. From the results, we can also see that appending multiple NMs consume much more decoding time than appending one NM. It is because NM is adopted for the chroma prediction when appending multiple NMs.

[22] also gave the encoding and decoding complexity increase compared with the anchor, thus we directly use the measured results in the literature. We can see that compared with [22], we can achieve better coding efficiency by all the proposals except replacing low probable TMs. However, the complexity also becomes higher since more network layers and larger dimensions are adopted in our proposal. For [23], we measure the encoding/decoding complexity for using one NM. About 14.2x and 132.3x higher encoding and decoding time is consumed compared with the anchor. Compared with [23], proposed appending one scheme can achieve better coding efficiency with smaller encoding/decoding time. It is because [23] directly perform NM for the chroma prediction if the best luma mode is NM, which will increase the burden of the chroma prediction.

*D. Visualized Result of Prediction Blocks*

As shown in Section V.A, by using NMs, the coding efficiency can be improved in terms of an objective measurement (i.e. BD-rate). In order to further confirm the efficiency of NMs, we present the visualized results of four cases in Fig. 12. The red box is the raw block, and five surrounding blocks are the neighboring reference blocks. The predicted block by the specific NM and TM is illustrated in the figure.

For the case (a), two predicted blocks are generated by the NM1 and TM0 (Planar), respectively. We can see that there are several black regions in the top of the raw block, while this black region does not appear in the predicted block by TM0. It is because TM0 utilizes an interpolation-like prediction which will smooth the predicted pixels. As a result, some local textures will be lost. However, by using NM1, we can see that the local upper black region can be remained.

For the case (b), two predicted blocks are generated by the NM3-NA and TM0, respectively. In the raw block, most pixels are dark except the left-bottom regions. By using TM0, the upper half part in the predicted block is lighter than that in the raw block. However, by using NM3-NA, the upper half part of the predicted block is as dark as the raw block. Moreover, for the left-bottom region, the predicted pixels as light as the ones in the raw block.

For the case (c), two predicted blocks are generated by the NM3-HOR and TM11, respectively. In the raw block, for the bottom parts, there are several dark horizontal strips. Therefore, the best TM is 11. For the predicted blocks of NM3-HOR, we can see that we can not only achieve a good prediction for the bottom dark horizontal strips, but the other parts are also close to the raw block.

For the case (d), two predicted blocks are generated by the NM3-VER and TM27, respectively. In the raw block, for the

This article has been accepted for publication in a future issue of this journal, but has not been fully edited. Content may change prior to final publication. Citation information: DOI 10.1109/TMM.2019.2963620, IEEE Transactions on Multimedia

1520-9210 (c) 2019 IEEE. Personal use is permitted, but republication/redistribution requires IEEE permission. See http://www.ieee.org/publications_standards/publications/rights/index.html for more information.
Authorized licensed use limited to: WASEDA UNIVERSITY LIBRARY. Downloaded on May 06,2020 at 16:55:33 UTC from IEEE Xplore. Restrictions apply.

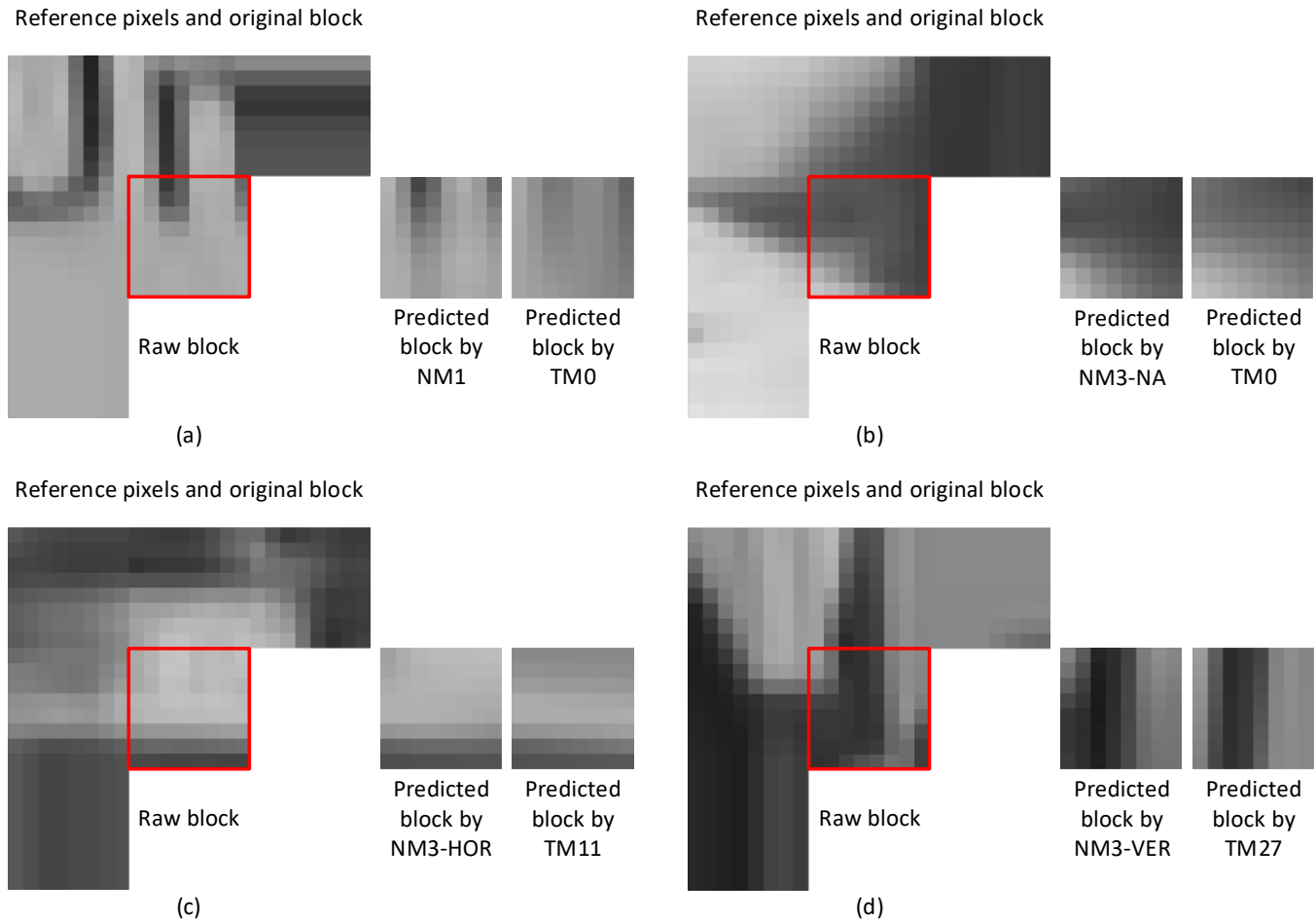

Fig. 12 Visualized results for best NM and best TM prediction.

vertical middle part, there are several dark vertical strips. Therefore, the best TM is 27. For the predicted blocks of NM3-VER, we can see that we can not only achieve a good prediction for the middle dark vertical strips, but the other parts are also close to the raw block.

*E. Analysis on Training QP*

We have analyzed the influence of this parameter (training QP) by doing the following three experiments. (1) set QP={22,27,32,37} for the training to evaluate the effect of using multiple QPs. (2) set QP=37 for the training to evaluate the effect of using large QP. (3) set QP=22 for the training to evaluate the effect of using another moderate QP.

For all the three experiments, the batch size and the total iteration is the same as the case of QP being 27. When applying the trained models in the appending scheme with one NM, the BD-rate compared with the anchor is shown in Table VIII.

When using multiple QPs (22, 27, 32, 37) as the training set, the average BD-rate is -4.5%, -6.1% and -6.7% for three channels respectively. When using QP22, the average BD-rate is -4.3%, -6.1% and -6.7%. When using QP37, the average BD-rate is -3.4%, -4.0% and -4.2%. We can see that using multiple QPs or using another moderate QP will not influence the final coding results so much. However, if we use a large QP such as 37, the coding gain will decrease obviously. It is because that the reconstructed pixels of large QP have poor quality, thus the learned mapping from the reference reconstructed blocks to the current blocks cannot fit an optimal regression in most scenarios.

*F. Ablation Study*

We have conducted the ablation study in terms of the number of used networks, the number of layers and the number of nodes in each layer. For each case, we evaluate the BD-rate saving, the encoding and decoding complexity. The ablation study of appending one NM is shown in Table IX. From the results, we can see that when increasing the number of layers from three to four, more coding gain can be achieved. For example, when using 512 nodes, the average BD-rate saving is 4.0%, 5.6% and 5.8% for three layers. When increasing to four layers, 0.4%, 0.4% and 0.9% more BD-rate can be saved. However, when further increasing the number of layers from four to five, there is no obvious coding gain while the encoding/decoding complexity becomes higher. Therefore, we think that using four layers is a reasonable trade-off between coding gain and complexity.

TABLE VIII
CODING GAIN OF USING DIFFERENT QP FOR THE TRAINING

| Class | Sequence | QP27 Y | QP27 U | QP27 V | QP22 Y | QP22 U | QP22 V | QP37 Y | QP37 U | QP37 V | 4 QPs (22,27,32,37) Y | 4 QPs U | 4 QPs V |
|---|---|---|---|---|---|---|---|---|---|---|---|---|---|
| A1 (4K) | Tango | -15.6 | -21.5 | -17.7 | -13.4 | -20.5 | -16.3 | -12.5 | -13.8 | -10.4 | -16.4 | -23.0 | -18.6 |
| | Drums100 | -3.7 | -5.0 | -5.7 | -4.0 | -5.4 | -6.4 | -3.3 | -4.0 | -5.3 | -4.0 | -5.8 | -6.8 |
| | CampfireParty | -3.6 | -10.3 | -12.9 | -3.4 | -11.1 | -13.9 | -2.1 | -6.1 | -6.0 | -4.2 | -10.3 | -14.3 |
| | ToddlerFountain | -2.8 | -4.8 | -2.1 | -3.0 | -4.8 | -2.1 | -2.5 | -6.1 | -2.4 | -3.2 | -5.8 | -2.7 |
| **Average of Class A1** | | **-6.4** | **-10.4** | **-9.6** | **-5.9** | **-10.5** | **-9.7** | **-5.1** | **-7.5** | **-6.0** | **-6.9** | **-11.2** | **-10.6** |
| A2 (4K) | CatRobot | -5.5 | -9.9 | -8.4 | -4.8 | -9.4 | -8.9 | -3.5 | -4.8 | -4.4 | -5.5 | -9.1 | -8.6 |
| | TrafficFlow | -8.7 | -9.1 | -15.0 | -8.5 | -11.3 | -17.9 | -7.9 | -5.0 | -12.3 | -10.0 | -9.0 | -16.6 |
| | DaylightRoad | -8.3 | -18.7 | -16.1 | -8.2 | -17.7 | -16.4 | -5.4 | -10.9 | -7.9 | -8.1 | -14.7 | -15.7 |
| | Rollercoaster | -14.9 | -19.9 | -13.9 | -13.0 | -16.7 | -12.1 | -8.7 | -8.7 | -5.4 | -14.0 | -19.1 | -11.1 |
| **Average of Class A2** | | **-9.3** | **-14.4** | **-13.3** | **-8.6** | **-13.8** | **-13.8** | **-6.4** | **-7.4** | **-7.5** | **-9.4** | **-13.0** | **-13.0** |
| A (WQXGA) | Traffic | -3.1 | -3.7 | -5.7 | -3.2 | -4.5 | -5.8 | -2.9 | -3.5 | -5.2 | -3.2 | -4.4 | -5.6 |
| | PeopleOnStreet | -3.2 | -2.7 | -5.5 | -3.5 | -3.8 | -5.8 | -3.1 | -3.4 | -5.3 | -3.6 | -3.0 | -6.7 |
| | Nebuta | -1.6 | -1.9 | -1.9 | -1.7 | -1.9 | -2.2 | -1.6 | -1.9 | -1.9 | -1.7 | -1.8 | -2.0 |
| | SteamLocomotive | -2.9 | -7.7 | -3.5 | -3.0 | -7.3 | -3.9 | -2.8 | -8.9 | -4.8 | -3.1 | -8.8 | -3.0 |
| **Average of Class A** | | **-2.7** | **-4.0** | **-4.1** | **-2.8** | **-4.4** | **-4.4** | **-2.6** | **-4.4** | **-4.3** | **-2.9** | **-4.5** | **-4.3** |
| B (1080P) | Kimono | -6.0 | -5.8 | -7.9 | -5.8 | -6.3 | -8.3 | -5.3 | -5.3 | -6.9 | -6.7 | -7.2 | -9.4 |
| | ParkScene | -2.2 | -4.9 | -2.9 | -2.4 | -3.7 | -3.0 | -2.0 | -1.8 | -1.1 | -2.7 | -3.4 | -3.0 |
| | Cactus | -2.3 | -2.9 | -5.8 | -2.5 | -2.6 | -4.7 | -1.9 | -1.7 | -4.0 | -2.6 | -2.4 | -6.2 |
| | BQTerrace | -9.7 | -13.2 | -9.5 | -9.6 | -13.2 | -12.6 | -6.7 | -8.3 | -8.1 | -8.2 | -10.4 | -8.9 |
| | BasketballDrive | -2.4 | -4.8 | -8.3 | -2.6 | -5.7 | -6.0 | -1.8 | -4.1 | -0.3 | -2.0 | -0.7 | -8.3 |
| **Average of Class B** | | **-4.5** | **-6.3** | **-6.8** | **-4.6** | **-6.3** | **-6.9** | **-3.5** | **-4.2** | **-4.1** | **-4.4** | **-4.8** | **-7.2** |
| C (WVGA) | BasketballDrill | 0.0 | -0.8 | -0.7 | 0.1 | -1.7 | -1.3 | -0.1 | -0.2 | -0.3 | 0.0 | -1.3 | -0.7 |
| | BQMall | -1.6 | -1.0 | -3.8 | -1.8 | -2.1 | -3.1 | -1.3 | -0.2 | -1.8 | -1.5 | 0.3 | -2.8 |
| | PartyScene | -1.1 | -2.0 | -1.5 | -1.3 | -1.6 | -1.4 | -1.0 | -1.3 | -0.5 | -1.1 | -1.5 | -1.3 |
| | RaceHorsesC | -1.4 | -1.5 | -1.3 | -1.4 | -1.2 | -1.9 | -1.4 | -1.9 | -1.2 | -1.5 | -1.9 | -1.6 |
| **Average of Class C** | | **-1.0** | **-1.3** | **-1.8** | **-1.1** | **-1.7** | **-1.9** | **-1.0** | **-0.9** | **-1.0** | **-1.0** | **-1.1** | **-1.6** |
| D (WQVGA) | BasketballPass | -1.3 | -3.9 | -3.0 | -0.8 | -2.4 | -3.5 | -1.5 | -4.4 | -1.2 | -1.3 | -3.6 | -0.6 |
| | BQSquare | -0.9 | 2.3 | -3.9 | -0.9 | 3.9 | -4.1 | -0.8 | 2.8 | -7.2 | -1.1 | -0.9 | -4.6 |
| | BlowingBubbles | -1.1 | -1.4 | -0.3 | -1.2 | -1.8 | 0.4 | -0.8 | -1.1 | -0.5 | -1.0 | -0.7 | -1.0 |
| | RaceHorses | -1.3 | -1.1 | -1.6 | -1.6 | -0.1 | -1.2 | -1.1 | -0.4 | -0.6 | -1.2 | -1.4 | -0.7 |
| **Average of Class D** | | **-1.2** | **-1.0** | **-2.2** | **-1.1** | **-0.1** | **-2.1** | **-1.0** | **-0.7** | **-2.4** | **-1.2** | **-1.7** | **-1.7** |
| E (720P) | FourPeople | -4.7 | -5.4 | -7.0 | -4.9 | -3.5 | -6.9 | -3.5 | -1.8 | -3.7 | -4.5 | -2.7 | -7.4 |
| | Johnny | -7.7 | -8.1 | -9.9 | -8.1 | -8.6 | -8.3 | -5.1 | -5.8 | -2.0 | -7.8 | -12.1 | -9.0 |
| | KristenAndSara | -5.8 | -3.2 | -9.5 | -6.1 | -5.0 | -9.1 | -3.8 | -0.1 | -6.6 | -5.8 | -5.0 | -10.5 |
| **Average of Class E** | | **-6.1** | **-5.6** | **-8.8** | **-6.4** | **-5.7** | **-8.1** | **-4.1** | **-2.6** | **-4.1** | **-6.0** | **-6.6** | **-9.0** |
| **Average** | | **-4.4** | **-6.2** | **-6.6** | **-4.3** | **-6.1** | **-6.7** | **-3.4** | **-4.0** | **-4.2** | **-4.5** | **-6.1** | **-6.7** |

TABLE IX
ABLATION STUDY OF APPENDING ONE NM

| | 3 layer BD-rate Y | U | V | Complexity Enc | Dec | 4 layer BD-rate Y | U | V | Complexity Enc | Dec | 5 layer BD-rate Y | U | V | Complexity Enc | Dec |
|---|---|---|---|---|---|---|---|---|---|---|---|---|---|---|---|
| node | | | | | | | | | | | | | | | |
| 512 | -4.0 | -5.6 | -5.8 | 246 | 1025 | -4.4 | -6.0 | -6.7 | 285 | 1378 | -4.3 | -6.2 | -6.6 | 331 | 1709 |
| 1024 | -4.0 | -5.7 | -6.2 | 359 | 1936 | -4.4 | -6.2 | -6.6 | 566 | 3821 | -4.7 | -6.2 | -7.0 | 908 | 6782 |
| 2048 | -3.9 | -5.7 | -6.2 | 1242 | 9406 | -4.4 | -6.2 | -6.8 | 2376 | 19690 | -4.5 | -6.4 | -6.7 | 3362 | 28264 |

In the case of four layers, we also conduct the ablation study of appending three NMs, as shown in Table X. When increasing the number of nodes from 512 to 1024, more coding gain can be achieved especially for the U and V channels. However, if we further increase the number of nodes from 1024 to 2048, there is no obvious coding gain, while the complexity becomes much higher. Therefore, we think that using 1024 nodes is a reasonable trade-off between coding gain and complexity.

TABLE X
ABLATION STUDY OF APPENDING THREE NMS WITH FOUR LAYERS

| node | Y-BD (%) | U-BD (%) | V-BD (%) | Enc (%) | Dec (%) |
|---|---|---|---|---|---|
| 512 | -5.5 | -6.2 | -7.3 | 992 | 4485 |
| 1024 | -5.6 | -7.2 | -7.9 | 3149 | 16860 |
| 2048 | -6.0 | -7.0 | -8.0 | 10903 | 65196 |

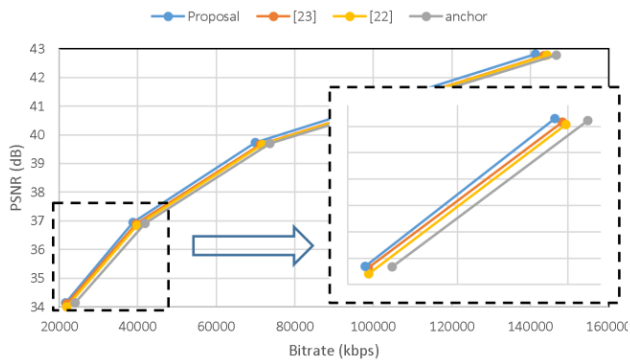

Fig. 13 R-D curves of the proposal, previous works and anchor. The result is for the average results of all the test sequences under four test QP22, 27, 32 and 37.

*G. R-D Curve and Subjective Results*

We have added the R-D curves for the proposed appending scheme with 7 NMs, the compared methods [22]-[23] and the original HEVC in Fig. 13. The result is for the average results of all the test sequences under four test QP22, 27, 32 and 37. We can see that compared with the anchor and the previous works, our proposal's curve locates at the left and top region, which indicates that the proposal can reach small bitrates or high PSNR.

In addition, we also evaluate the R-D curve of six high QPs (QP37~QP42) and subjective results for two test sequences Kimono and KristenAndSara as shown in Fig. 14-17. From the R-D curves, we can see that by using our methods, there is a significant coding gain for both sequences. The subjective results for the two sequences are shown in Fig. 16 and Fig. 17, respectively. For Kimono, when using our proposal, we can obtain better subjective results especially for the human face. For KristenAndSara, when using our proposal, we can obtain better subjective results especially for the sleeve.

## VI. CONCLUSIONS

In this paper, we develop an intra prediction method driven by multiple NMs. Two types of strategies are presented to integrate NMs with TMs. Compared with the previous work, when appending seven NMs, we can save 2.6%, 3.8%, 3.1% BD-rates for Y, U, V channels respectively. Besides, we have also visualized the predicted blocks by NMs to verify the effectiveness.

About the future work, since the effect of multiple NMs has been verified for a fixed block size 8×8, thus we will extend to variable block sizes. Second, as shown in Section V.C, the encoding/decoding complexity is much higher than the anchor without using NMs. Therefore, the acceleration for the NM processing is highly required.

## ACKNOWLEDGMENT

The authors would like to thank Dr. Jiahao Li from Peking University for his great helps on this work.

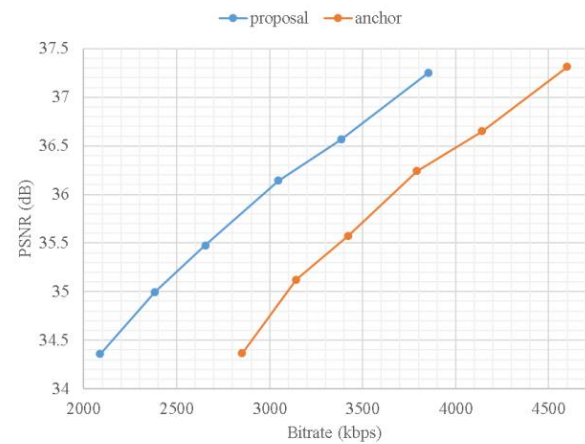

Fig. 14 R-D curves of the proposal and anchor. The result is for the sequence *Kimono* under QP37-QP42.

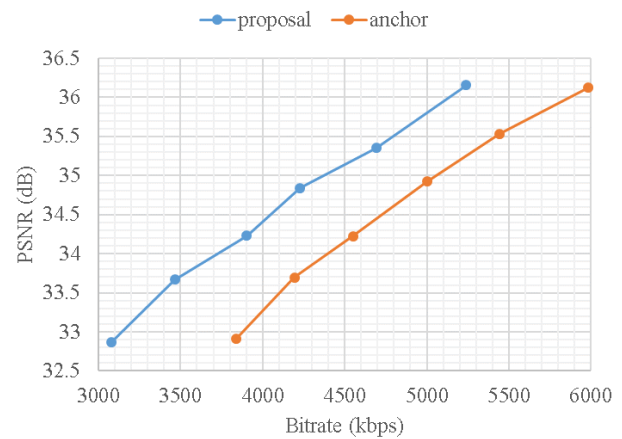

Fig. 15 R-D curves of the proposal and anchor. The result is for the sequence *KristenAndSara* under QP37-QP42.

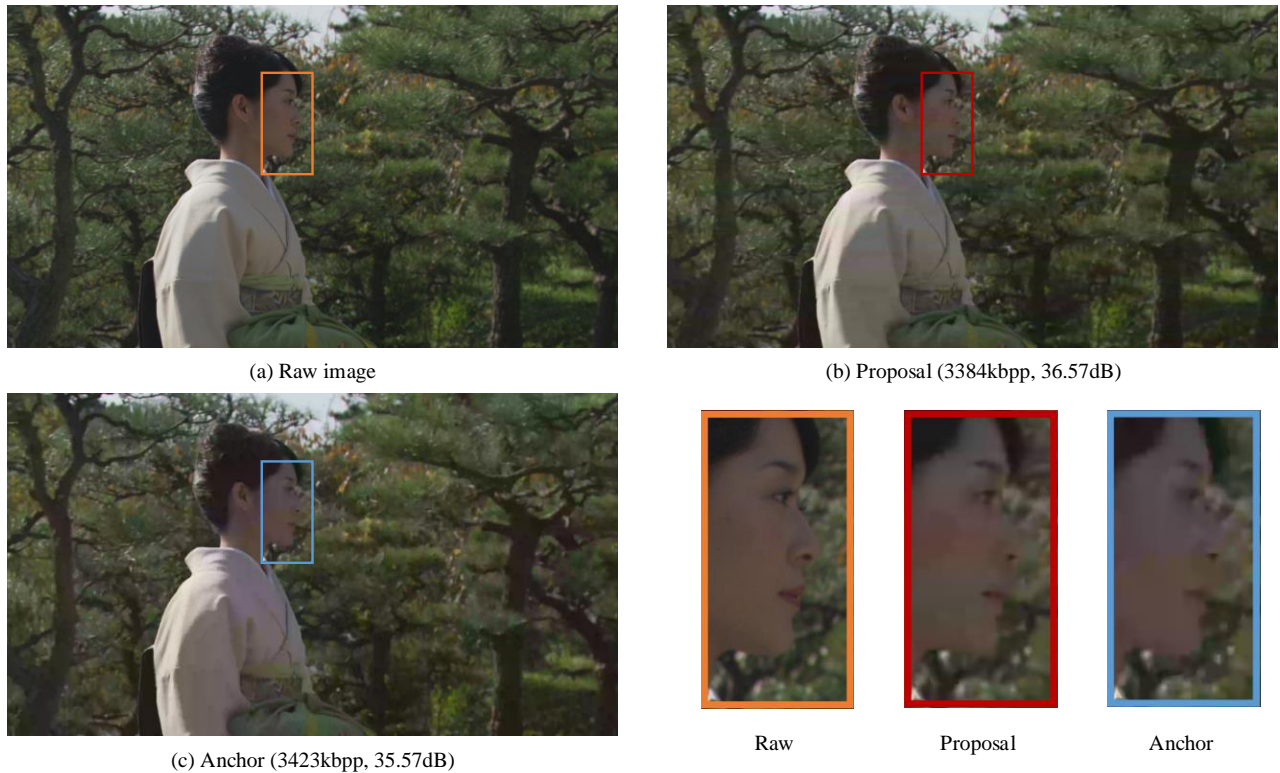

Fig. 16 Subjective results of the proposal and anchor for the sequence *Kimono*.

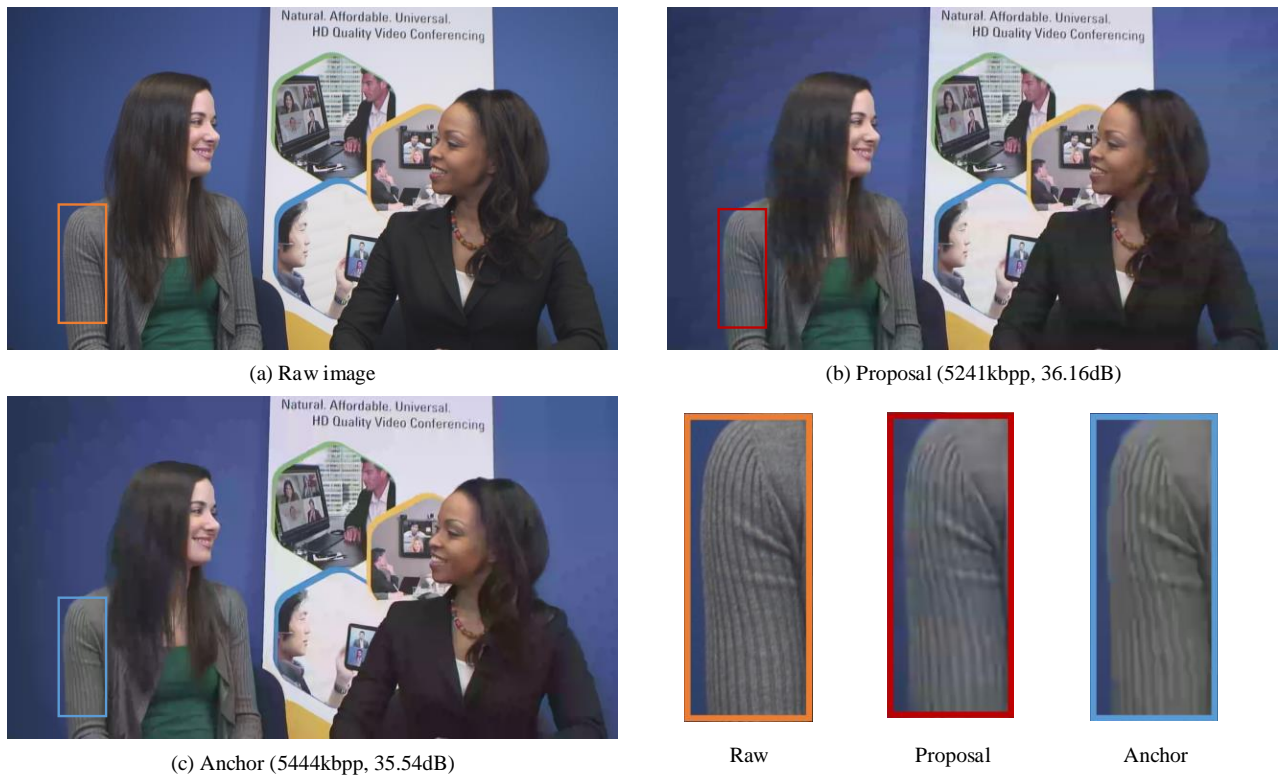

Fig. 17 Subjective results of the proposal and anchor for the sequence *KristenAndSara*.

Authors' photographs and biographies not available at the time of publication.